\documentclass[11pt,a4paper,final]{article}

\usepackage[a4paper,left=2.8cm,right=2.8cm,top=3cm,bottom=3cm]{geometry}

\usepackage[utf8]{inputenc}
\usepackage[USenglish]{babel}

\usepackage{amsmath}
\usepackage{amsfonts}
\usepackage{amssymb}
\usepackage{graphicx}
\usepackage{xcolor}
\usepackage{setspace}
\usepackage{hyperref}
\usepackage{siunitx}

\usepackage[width=0.95\linewidth]{caption}
\usepackage[list=true]{subcaption}

\newcommand{\dext}{\text{d}}
\newcommand{\GN}{G_\text{N}}

\newcommand{\ms}{\ensuremath{m_\text{s}}}

\newcommand{\tp}{\ensuremath{t_\text{p}}}

\usepackage{hyperref}
\hypersetup{
	pdftitle={Early-Stage Shear Viscosity far from Equilibrium via Holography},
	pdfauthor={Michael Florian Wondrak, Matthias Kaminski, Marcus Bleicher},
	pdffitwindow=true,
	colorlinks=true,
	linkcolor={red!50!black},
	citecolor={blue!50!black},
	urlcolor={blue!80!black}
}

\usepackage[comma,numbers,sort&compress]{natbib}
\begin{document}

\begin{center}
\begin{spacing}{2}
{\Large\bf Early-Stage Shear Viscosity far from Equilibrium\\via Holography}
\end{spacing}
\end{center}

\vspace{-1cm}
\begin{center}
\textbf{Michael~F.~Wondrak}$^{a,}$\footnote{Corresponding author, \texttt{wondrak@itp.uni-frankfurt.de}},
\textbf{Matthias~Kaminski}$^{b,}$\footnote{\texttt{mski@ua.edu}},
\textbf{and Marcus~Bleicher}$^{a,}$\footnote{\texttt{bleicher@th.physik.uni-frankfurt.de}}

\vspace{.6truecm}
{\em $^a$Institut f\"ur Theoretische Physik,\\
Johann Wolfgang Goethe-Universit\"at Frankfurt am Main,\\
Max-von-Laue-Stra\ss{}e 1, 60438 Frankfurt am Main, Germany}\\

\vspace{.3truecm}
{\em $^b$Department of Physics and Astronomy, University of Alabama, \\
514 University Blvd., Tuscaloosa, AL 35487, USA}\\

\vspace{.6truecm}
April 9, 2020
\end{center}

\vspace{0.1cm}

\begin{abstract}
\noindent{\small%
\noindent Shear viscosity is a crucial property of QCD matter which determines the collective behavior of the the quark-gluon plasma (QGP) in ultrarelativistic heavy-ion collisions.
Extending the near-equilibrium, high-precision investigations in theory and experiment, we take into account the fact that, in a collision, the QGP is generated far from equilibrium.
We use the AdS/CFT correspondence to study a strongly coupled plasma and find a significant impact on the ratio of shear viscosity to entropy density, $\eta/s$. 
In particular, we investigate the initial heating phase and find a decrease reaching down to below $60\%$ followed by an overshoot to $110\%$ of the near-equilibrium value.
This finding might be highly relevant for the extraction of transport coefficients from anisotropic flow measurements at RHIC and LHC.

}
\end{abstract}

\clearpage


\section{Introduction}
\label{sec:Introduction}
Heavy-ion collisions are violent processes: Two nuclei collide, heat up above the critical temperature $T_\text{C} \approx \SI{155}{MeV}$, and produce a fireball of deconfined quarks and gluons. This quark-gluon plasma (QGP) finally hadronizes and emits particles which can be measured in particle detectors. Despite the fact that the QGP exists only for some $\text{fm}/\text{c}$, advances in theory and experiment allow to study its properties at high precision. In particular, transport coefficients are under scrutiny. A prime example is the shear viscosity, $\eta$, which is the dominating factor for elliptic flow,~$v_2$, of charged particles. 

It was the AdS/CFT correspondence which predicted a small and constant value of the shear viscosity for a large class of strongly-coupled gauge theories. When expressed in terms of the entropy density, the value reads $\eta/s = 1/4\pi$ when expressed in natural units ($\text{c}\equiv\hbar\equiv k_\text{B}\equiv 1$) \cite{Kovtun:2004de}.
Hydrodynamic simulations extract similar values from experimental flow data \cite{Gyulassy:2004zy,Romatschke:2007mq,%
Schenke:2011tv,Acharya:2020taj}.
The theoretical calculation of $\eta/s$ is challenging because the QGP is non-dilute and strongly coupled. 
Most notable are the temperature-dependent results by lattice QCD \cite{Astrakhantsev:2017nrs} and the functional renormalization group (FRG) \cite{Christiansen:2014ypa}.
However, these approaches are restricted to the near-equilibrium regime.
We aim at overcoming this limitation and improving the description of the early collision phase~\cite{Wondrak:2020tzt} which is far from equilibrium~\cite{Romatschke:2017vte}. 
We apply the AdS/CFT correspondence and focus on the initial heating phase.

\section{Holographic Setup}
\label{sec:Holographic_Setup}
\begin{figure}[!ht]
  \centering
  \includegraphics[width=0.9\linewidth]{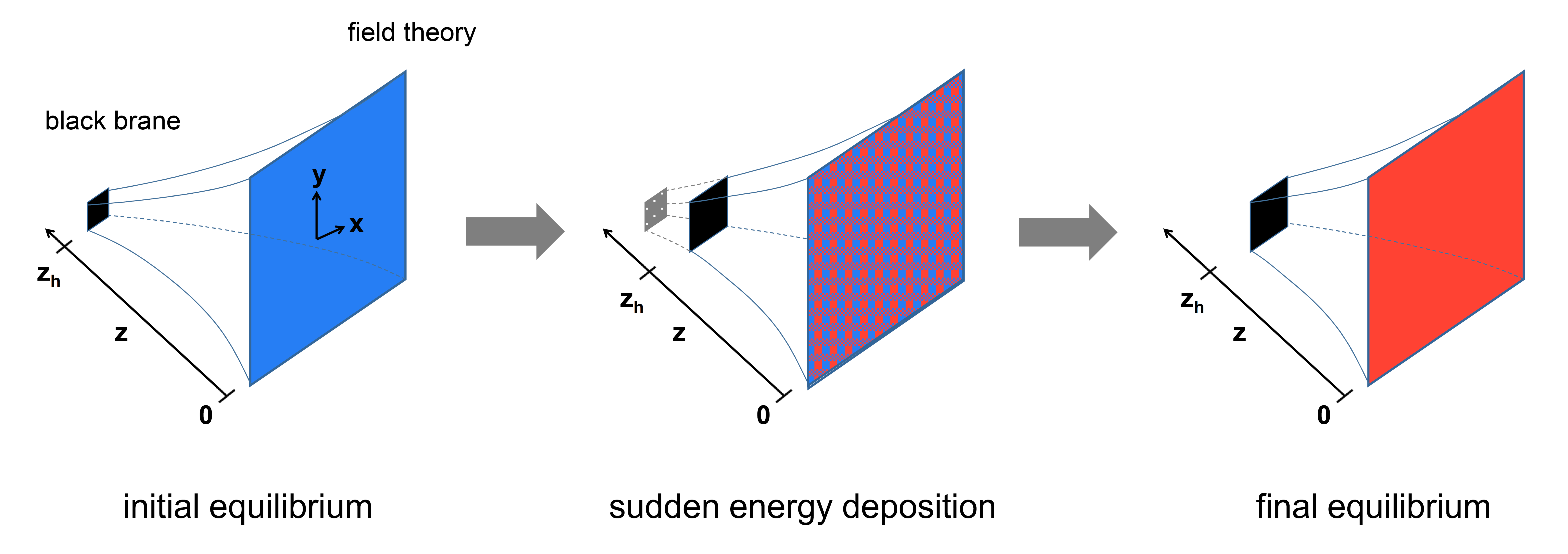}
  \caption{Sketch of the holographic system during the time evolution through the far from equilibrium regime. The coordinate $z$ denotes the inverse radial coordinate, the black brane horizon is located at~$z_\text{h}$. The rapid energy deposition in the boundary theory corresponds to a rapid mass increase of the black brane. 
  }
  \label{fig:Fig1}
\end{figure}

The appeal of the AdS/CFT correspondence in the field of heavy-ion physics is the equivalence of a thermal field theory state to a black brane configuration, a black hole with planar horizon. 
Technically speaking, the AdS/CFT correspondence identifies certain pairs of strongly coupled gauge theories at large gauge group rank and supergravity on spacetimes with Anti-de Sitter asymptotics. The gravitational spacetime is referred to as ``bulk'' while the field theory lives on its ``boundary'': Introducing the inverse radial direction $z$, the plasma is located at $z=0$, the black brane horizon lies at $z_\text{h}$, and the singularity is found at $z \to \infty$, cf.~Fig.~\ref{fig:Fig1}. 
The thermodynamic state variables of plasma and black brane are closely related. In equilibrium, the field-theory temperature equals the black brane Hawking temperature, and the entropy agrees with the Bekenstein-Hawking entropy.

On the field-theory side, we extend the calculation of $\eta/s$ of a strongly coupled plasma to the far-from-equilibrium regime. We consider a sudden homogeneous energy deposition as a model for the heating phase during a heavy-ion collision.
On the gravity side, this is realized by a rapid mass accretion onto a black brane. We use a Reissner-Nordstr\"om Vaidya black brane which is a solution of Einstein's general relativity coupled to Maxwell theory \cite{Vaidya:1951zz,Caceres:2012em,Wondrak:2017kgp}.
The line element in infalling Eddington-Finkelstein coordinates reads
\begin{equation}
ds^2 
= \frac{1}{z^2} 
 \left( -f(v, z)\, \dext v^2 -2\, \dext v\, \dext z  +\dext x^2 +\dext y^2 \right).
\end{equation}
The coordinate $v$ denotes the ingoing null time and agrees with the ordinary field-theory time $t$ at the boundary, $z=0$. The blackening factor
$
f(v,z)
=1 -2\GN M(v)\, z^3 +\GN Q(v)^2\, z^4
$
depends on the black brane's mass and charge parameters, $M(v)$ and $Q(v)$, and determines the horizon position.
The rapid mass infall is given as
\begin{equation}
M(v)
= m +\ms\; \left(1 +\tanh\left(v/\Delta t\right)\right)/2.
\end{equation}

The extension of the thermodynamic variables to the time-dependent regime is non-trivial. We use the grand potential, provided by the AdS/CFT correspondence, to define the field-theory temperature and entropy density, $T(t)$ and $s(t)$, in terms of gravitational degrees of freedom.

\section{Spacetime Perturbations and \texorpdfstring{$\eta/s$}{eta/s}}
\label{sec:Spacetime_Perturbations}
According to the AdS/CFT correspondence, the bulk metric is dual to the energy-momentum tensor of the field-theory plasma. In particular, the ring-down of a geometry perturbation $h_{mn}$ yields the evolution of the expectation value $\left\langle T^{\mu\nu}\!\left(t\right) \right\rangle_{h}$.
Applying linear response theory, we obtain the retarded Green's function, $G_\text{R}^{xy,xy}(\tp,\, t_2)$, from the 1-point function $\left\langle T^{xy}\!\left(t_2\right) \right\rangle_{h}$ if the perturbation is localized at a time $\tp$:
\begin{equation}
\left\langle T^{xy}\!\left(t_2\right) \right\rangle_{h}
= \int\!\dext \tau\; G_\text{R}^{xy,xy}\!\left(\tau,\, t_2\right) \, \underbrace{h^{(0)}_{xy}\!\left(\tau\right)}_%
{\propto\, \delta\left(\tau -t_\text{p}\right)} 
\propto G_\text{R}^{xy,xy}(t_\text{p},\,t_2)
\end{equation}

\begin{figure}[!ht]%
\begin{subfigure}[h]{0.45\textwidth}
  \includegraphics[width=\textwidth]{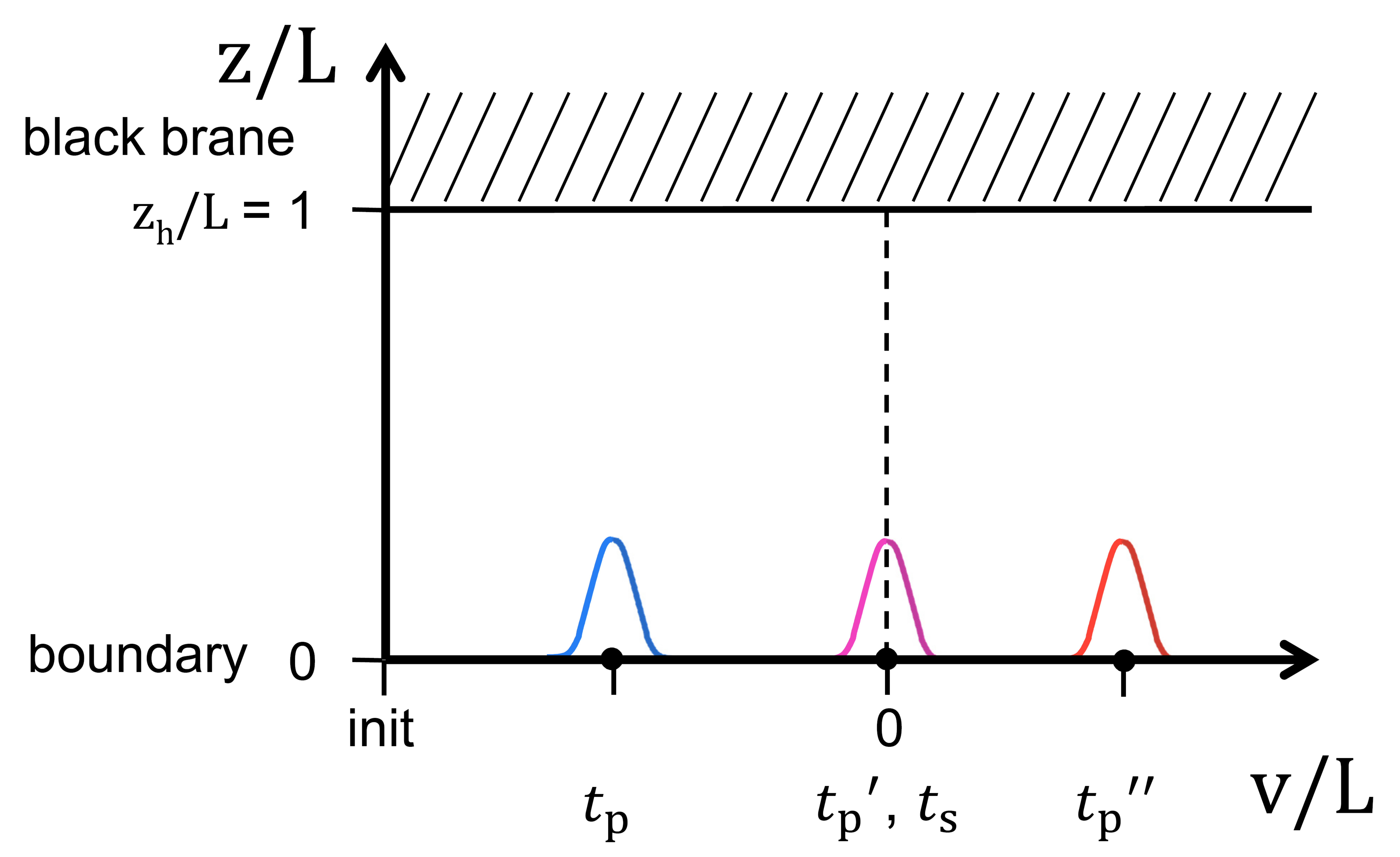}
  \caption{Sketch of the bulk spacetime including the metric perturbations~$h^\text{(0)}_{xy}$. Sample perturbations are introduced at times $t_\text{p}$ before, at the same time, or after the energy deposition at $t_\text{s}=0$. Note that a redefinition of the coordinate $z$ compensates the shift in the horizon position, $z_\text{h}$.}
  \label{fig:Fig2a}
\end{subfigure}
\hfill
\begin{subfigure}[h]{0.45\textwidth}
  \includegraphics[width=\linewidth]{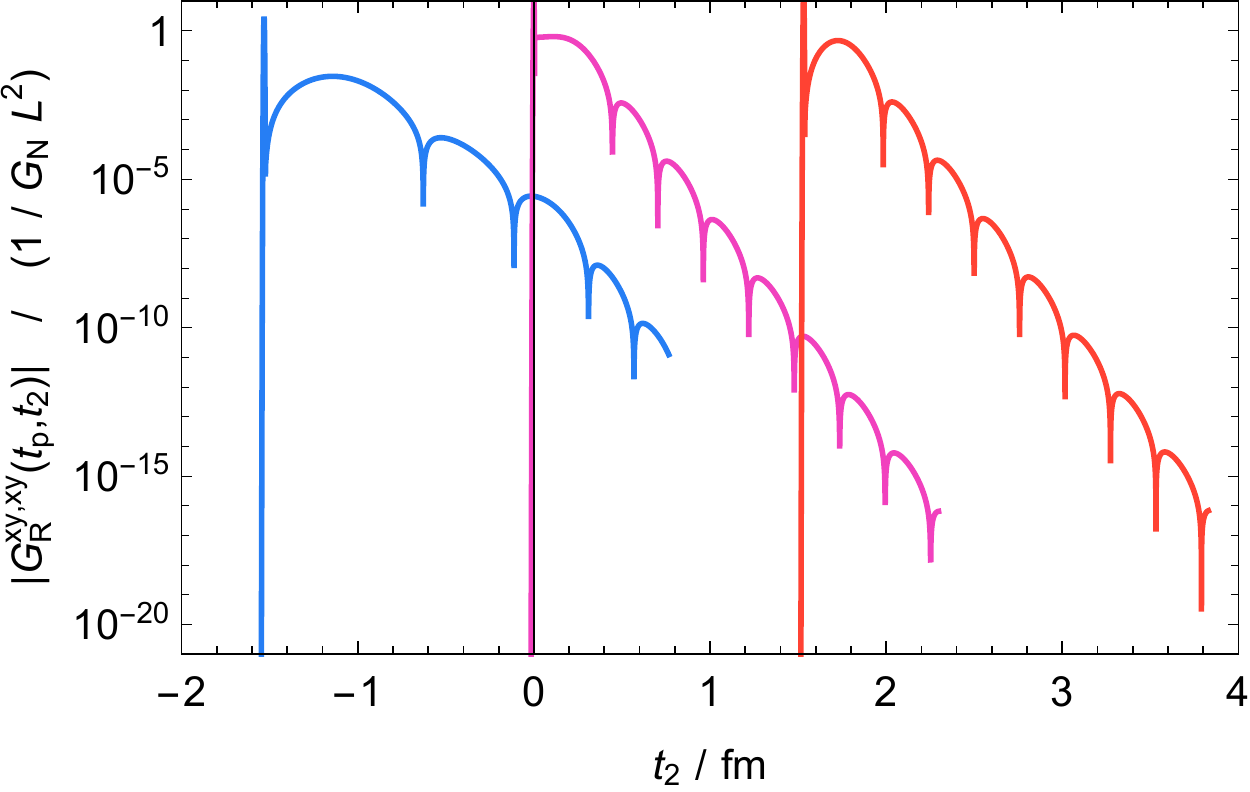}
  \caption{The retarded Green's function in configuration space, $G_\text{R}^{xy,xy}(t_\text{p},\, t_2)$, shows damped oscillations: the dominant quasi-normal mode. Note that the period and damping scale with temperature, which is increased by a factor of two at $t_2=0$ (cf.~Fig.~\ref{fig:Fig3}).%
}
  \label{fig:Fig2b}
\end{subfigure}%
\caption[Gret]{Introducing perturbations to extract the shear correlator. $L$~denotes a gravitational length scale. The colors indicate at which regime of the background evolution the perturbation is introduced, cf.~Fig~\ref{fig:Fig1}.}
\label{fig:Fig2}
\end{figure}

The solution to the evolution of the metric fluctuation is found numerically~\cite{Chesler:2013lia,Banerjee:2016ray}. 
A pseudospectral method is used on the spatial part, while a fourth-order Runge-Kutta scheme evolves the system in time. Three samples are presented in Fig.~\ref{fig:Fig2}.

We arrive at a momentum-space version of the retarded Green's function, $\tilde{G}_\text{R}^{xy,xy}(t_\text{avg},\omega)$, by a Wigner transformation. It depends on the frequency $\omega$, which is the momentum-space conjugate of the relative time, $t_\text{p}-t_2$, and on the average time, $t_\text{avg} = (t_\text{p} +t_2)/2$ \cite{Balasubramanian:2012tu}. 
The time-dependent shear viscosity is defined by the time-dependent Kubo formula, 
\begin{equation}
\eta\,(t_\text{avg})
=-\lim\limits_{\omega \to 0}\, \frac{1}{\omega} \,\Im\, \tilde{G}_\text{R}^{xy,xy}(t_\text{avg},\omega).
\end{equation}

\begin{figure}[!ht]
  \centering
  \includegraphics[width=0.6\linewidth]{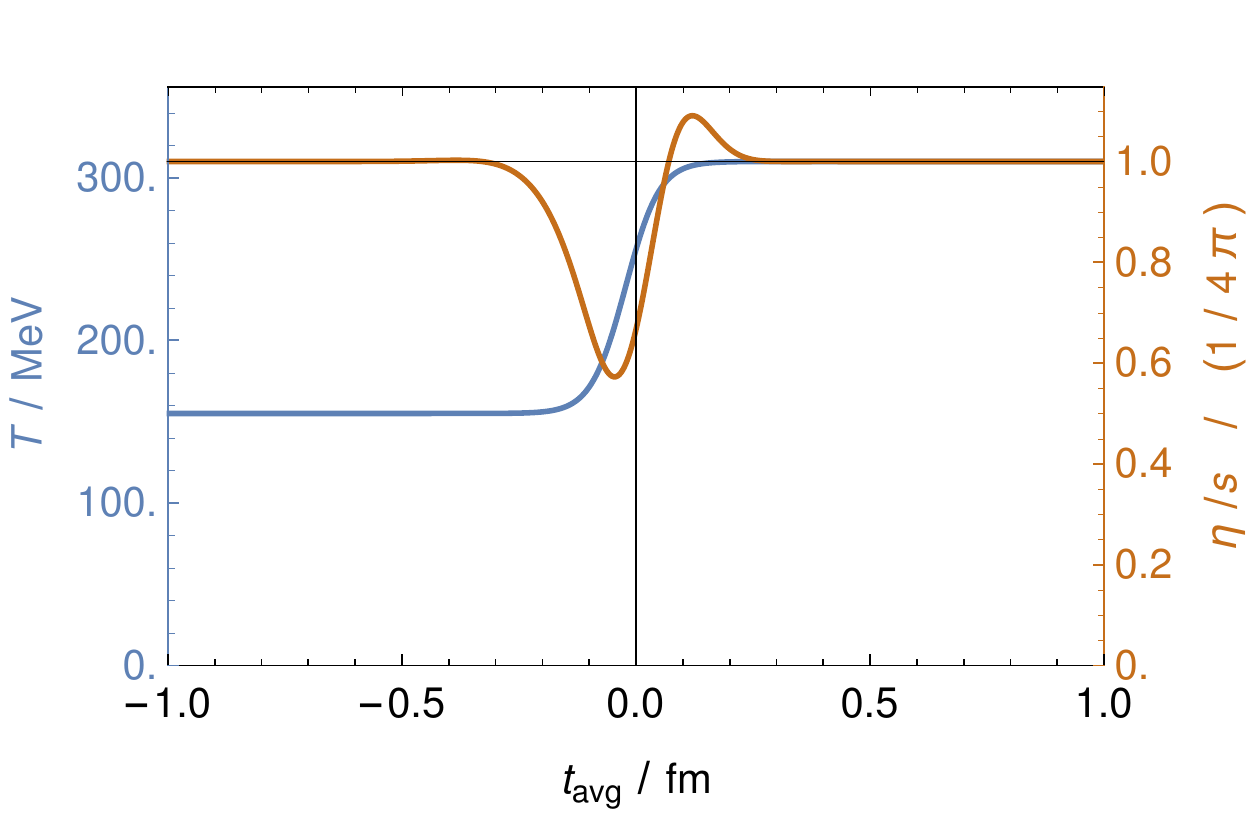}
  \captionof{figure}{Shear viscosity to entropy density ratio, $\eta/s$ (orange), far from equilibrium for a given temperature temperature profile (blue). The evolution shows significant corrections to the near-equilibrium value~$1/4\pi$.}
  \label{fig:Fig3}
\end{figure}

At RHIC, a temperature increase to $\SI{310}{MeV}$ within $\SI{0.3}{fm}$ is typical for the early phase of a collision at $\sqrt{s_\text{NN}}=\SI{200}{GeV}$ \cite{Adare:2009qk}. 
Figure~\ref{fig:Fig3} presents the corresponding evolution of $\eta/s$. Asymptotically, the curve takes the near-equilibrium value which amounts to $1/4\pi$ in our case~%
\footnote{A common misunderstanding should be pointed out: The number $1/4\pi$ is not a universal lower bound, neither for all quantum field theories with a gravity dual, nor for every quantum field theory at strong coupling. %
However, every near-equilibrium holographic model studied thus far has a lower bound, e.g.~Ref.~\cite{Brigante:2008gz}.%
}. 
During the far-from-equilibrium period, however, there are significant corrections: The value of $\eta/s$ reaches down to below 60\% and increases to 110\% of the near-equilibrium result. 

\section{Conclusions and Outlook}
\label{sec:Conclusions_outlook}
We presented the first holographic non-equilibrium calculation of $\eta/s$ via the retarded Green's function. Far from equilibrium, the viscosity-to-entropy ratio changes drastically. We expect comparable corrections to apply to the near-equilibrium results of FRG and lattice QCD. Our findings directly impact hydrodynamic simulations and the extraction of viscosity from experimental data.

\section*{Acknowledgments}
MFW expresses his thanks to the selection committee for the flash talk award.
We thank C.~Cartwright, T.~Ishii, and P.~Nicolini for discussions.
This work was supported, in part, by the U.S.~Department of Energy grant DE-SC-0012447, by the Helmholtz International Center for FAIR (HIC for FAIR) within the LOEWE program launched by the State of Hesse, by the Helmholtz Graduate School for Hadron and Ion Research (HGS-HIRe for FAIR), by the Stiftung Polytechnische Gesellschaft Frankfurt am Main, and by the Studienstiftung des deutschen Volkes.


\providecommand{\href}[2]{#2}
\begingroup\raggedright\endgroup


\begin{thebibliography}{10}

\bibitem{Kovtun:2004de}
P.~K. Kovtun, D.~T. Son, and A.~O. Starinets {\em Phys. Rev. Lett.} {\bfseries
  94} (2005) 111601,
\href{http://arxiv.org/abs/hep-th/0405231}{{\ttfamily arXiv:hep-th/0405231
  [hep-th]}}.

\bibitem{Gyulassy:2004zy}
M.~Gyulassy and L.~McLerran {\em Nucl. Phys.} {\bfseries A750} (2005) 30--63,
\href{http://arxiv.org/abs/nucl-th/0405013}{{\ttfamily arXiv:nucl-th/0405013
  [nucl-th]}}.

\bibitem{Romatschke:2007mq}
P.~Romatschke and U.~Romatschke {\em Phys. Rev. Lett.} {\bfseries 99} (2007)
  172301,
\href{http://arxiv.org/abs/0706.1522}{{\ttfamily arXiv:0706.1522 [nucl-th]}}.

\bibitem{Schenke:2011tv}
B.~Schenke, S.~Jeon, and C.~Gale {\em Phys. Lett.} {\bfseries B702} (2011)
  59--63,
\href{http://arxiv.org/abs/1102.0575}{{\ttfamily arXiv:1102.0575 [hep-ph]}}.

\bibitem{Acharya:2020taj}
{\bfseries ALICE} Collaboration, S.~Acharya {\em et~al.}
\href{http://arxiv.org/abs/2002.00633}{{\ttfamily arXiv:2002.00633 [nucl-ex]}}.

\bibitem{Astrakhantsev:2017nrs}
N.~Astrakhantsev, V.~Braguta, and A.~Kotov {\em J.~High Energy Phys.}
  {\bfseries 04} (2017) 101,
\href{http://arxiv.org/abs/1701.02266}{{\ttfamily arXiv:1701.02266 [hep-lat]}}.

\bibitem{Christiansen:2014ypa}
N.~Christiansen, M.~Haas, J.~M. Pawlowski, and N.~Strodthoff {\em Phys. Rev.
  Lett.} {\bfseries 115} (2015) 112002,
\href{http://arxiv.org/abs/1411.7986}{{\ttfamily arXiv:1411.7986 [hep-ph]}}.

\bibitem{Wondrak:2020tzt}
M.~F. Wondrak, M.~Kaminski, and M.~Bleicher
\href{http://arxiv.org/abs/2002.11730}{{\ttfamily arXiv:2002.11730 [hep-ph]}}.

\bibitem{Romatschke:2017vte}
P.~Romatschke {\em Phys. Rev. Lett.} {\bfseries 120} (2018) 012301,
\href{http://arxiv.org/abs/1704.08699}{{\ttfamily arXiv:1704.08699 [hep-th]}}.

\bibitem{Vaidya:1951zz}
P.~Vaidya
{\em Proc. Natl. Inst. Sci. India} {\bfseries A33} (1951) 264.

\bibitem{Caceres:2012em}
E.~Caceres and A.~Kundu {\em J.~High Energy Phys.} {\bfseries 09} (2012) 055,
\href{http://arxiv.org/abs/1205.2354}{{\ttfamily arXiv:1205.2354 [hep-th]}}.

\bibitem{Wondrak:2017kgp}
M.~F. Wondrak, M.~Kaminski, P.~Nicolini, and M.~Bleicher {\em J. Phys. Conf.
  Ser.} {\bfseries 942} (2017) 012020,
\href{http://arxiv.org/abs/1711.08835}{{\ttfamily arXiv:1711.08835 [hep-th]}}.

\bibitem{Chesler:2013lia}
P.~M. Chesler and L.~G. Yaffe {\em J.~High Energy Phys.} {\bfseries 07} (2014)
  086,
\href{http://arxiv.org/abs/1309.1439}{{\ttfamily arXiv:1309.1439 [hep-th]}}.

\bibitem{Banerjee:2016ray}
S.~Banerjee, T.~Ishii, L.~K. Joshi, A.~Mukhopadhyay, and P.~Ramadevi {\em
  J.~High Energy Phys.} {\bfseries 08} (2016) 048,
\href{http://arxiv.org/abs/1603.06935}{{\ttfamily arXiv:1603.06935 [hep-th]}}.

\bibitem{Balasubramanian:2012tu}
V.~Balasubramanian, A.~Bernamonti, B.~Craps, V.~Keranen, E.~Keski-Vakkuri,
  B.~M\"uller, L.~Thorlacius, and J.~Vanhoof {\em J.~High Energy Phys.}
  {\bfseries 04} (2013) 069,
\href{http://arxiv.org/abs/1212.6066}{{\ttfamily arXiv:1212.6066 [hep-th]}}.

\bibitem{Adare:2009qk}
{\bfseries PHENIX} Collaboration, A.~Adare {\em et~al.} {\em Phys. Rev.}
  {\bfseries C81} (2010) 034911,
\href{http://arxiv.org/abs/0912.0244}{{\ttfamily arXiv:0912.0244 [nucl-ex]}}.

\bibitem{Brigante:2008gz}
M.~Brigante, H.~Liu, R.~C. Myers, S.~Shenker, and S.~Yaida {\em Phys. Rev.
  Lett.} {\bfseries 100} (2008) 191601,
\href{http://arxiv.org/abs/0802.3318}{{\ttfamily arXiv:0802.3318 [hep-th]}}.

\end{thebibliography}
\end{document}